\documentstyle[aps,preprint]{revtex}
\begin{document}
\author{F. Di Tolla$^{1,2}$, A. Dal Corso$^{1,2}$, J. A. Torres$^{3}$, E. Tosatti$^{1,2,3}$}
\title{Electronic Properties of Ultra-Thin Aluminum  Nanowires}
\date{\today}
\address{1: International School for Advance Studies (SISSA), Trieste, Italy.\\
2: Istituto Nazionale per la Fisica della Materia (INFM), Unit\`a
di Trieste, Italy.\\ 3: Abdus Salam International Center for
Theoretical Physics (ICTP), Trieste, Italy. }

\maketitle
\begin{abstract}
We have carried out first principles electronic structure and
total energy calculations for a series of ultrathin aluminum
nanowires, based on structures obtained by relaxing the model
wires of Gulseren et al\cite{Gulseren98}. The number of conducting
channels is followed as the wires radius is increased. The results
suggest that pentagonal wires should be detectable, as the only
ones who can yield a channel number between 8 and 10.
\end{abstract}

\section{Introduction}

Conductance quantization in break junctions\cite{Krans} and
tip-surface contacts \cite{Agrait,Pascual,Olsen94} proves the
existence of extremely narrow constrictions between bulk
conductors. These constrictions can be seen as very short
nanowires, for which several models
\cite{Torres,Martin,Stafford97,Campers} and first principle
descriptions \cite{Lang95,Barnet97,Yannouleas98,Kobayashi99} can
be found in the literature. In proximity of surface melting, much
longer nanowires have been predicted in simulation\cite{Tomagnini}
and found experimentally \cite{Kuipers93} for Pb. The detailed
shape of these long nanowires has not been directly accessible,
but simulations showed that they could be extremely thin and
regular.

Very recently, Takayanagi and his group were able to
demonstrate\cite {Takayanagi} that very long and regular gold
wires of decreasing radius and varying structure can be
stabilized, suspended in vacuum between two tips. While a study
for gold, whose wire structure and properties are rather
substantially complicated by the presence of s-d interplay
\cite{Takeuchi89} and by surface reconstruction phenomena
\cite{Ercolessi88,Andreoni91}, we have in the meanwhile carried
out a study for a simple s-p metal, aluminum, whose idealized wire
behavior can to some extent serve as a prototypical case study.

Actually also data on Al wires have been reported
\cite{Krans,Scheer,Balicas}. Krans and coworkers\cite{Krans} used
mechanically controllable break junctions (MCB) to measure the
conductance of Al one-atom point contacts at $T=4.2$\,K and found
a quantized behaviour with clear plateaux for the smallest integer
multiples of $2e^2/h$. Using a similar setup, Scheer and
coworkers\cite{Scheer} were able to decompose the total
transmission into contributions from various channels. This showed
that, even in the case of a single quantum of conductivity, at
least three channels were {\em open} and contributed to the total
transport. Balicas et al.\cite{Balicas} found that histograms of
conductance measurements, taken at room temperature, show a
structure with features around small integer multiples (1,3,...)
of the quantum of conductivity $2e^2/h$.

In this work we present realistic Generalized Gradient Approximation
(GGA) electronic structure calculations of several wires, and discuss in particular
the interplay between structure and number of conducting channels,
by comparing our results with the predictions of two simple
jellium models: a {\em hard-wall} cylinder, and a {\em soft-wall}
one with parameters appropriate for the Al surface\cite{Martin}.

The rest of the paper is organized as follows. In Section
\ref{sec:method} we describe the procedures, methods, and
approximations adopted for the wire calculations, as well as the
physical quantities which will be monitored. Section
\ref{sec:results} presents the resulting electronic structures,
total energies, and number of channels for a set of ultra-thin
wires of increasing thickness, ranging from monoatomic to
hexagonal. In section \ref{sec:discuss} we discuss our results,
and possible future calculations aimed at including phenomena
beyond the present level of description.

\section{Method}
\label{sec:method}

All calculations presented here are carried out for straight
wires, using a supercell technique. The wires are infinite and
periodic along the $z$-direction, and periodically repeated as a
square lattice with a very large spacing in the $x-y$ plane.

We have considered altogether 7 different wires with increasing
cross section, namely: monoatomic, diatomic, triatomic, centered
pentagonal (staggered and eclipsed), and centered hexagonal
(staggered and eclipsed). The initial structure for the wires was
taken from Gulseren et al.'s glue modeling \cite {Gulseren98},
with the understanding that this should represent no more than a
suggestion (glue models are not sufficiently accurate in this thin
wire regime) for further ab-initio refinement. Of course this
procedure does not ensure an exhaustive enumeration of all
possible low-energy, or ''magic'' thin wire structures.
Nonetheless, it does turn out that the glue model nanowires are
close enough to well defined local energy minima, so that we
believe we have at least a subset of the nanowires which real Al,
given the proper experimental conditions, would form.

The effect of strong electron correlations, and of classical and
quantum fluctuations, all of which are of course crucial in a one
dimensional system, are not included here. In particular, we will
not include at this level the possibility of Peierls or
spin-Peierls dimerizations, of ferromagnetism, and/or Mott-Hubbard
antiferromagnetic insulator states. These well-known phenomena and
their mutual competition of course do require more sophisticated
approaches. The scope of the present calculations  is also to lay
the ground with a description of the basic, mean-field metallic
state state of the wire, upon which these further approaches could
be later built.

The electronic structure calculations were performed within the
GGA, using the Becke-Perdew exchange-correlation functional \cite{Becke}.
The electron-ion interaction was described by means of a
pseudopotential by Stumpf et al. \cite{Gonze}, and the
wavefunctions were expanded in planewaves up to a cutoff energy
$E_{cut}=12$ Ry.

Within this scheme, bulk Al was calculated to posses a lattice
spacing of 3.95 \AA\ (experimental 4.0496 \AA ), and a cohesive
energy of 3.74 eV/atom, (experimental 3.39 eV/atom). We also
calculated the Al$_2$ molecule, which we found correctly to be an
electronic triplet with a length of 2.657 \AA . For most
calculations the {\bf k}-point summation was carried out directly
on a regular mesh of 10 {\bf k}-points in half of the one
dimensional BZ along $k_z$, while only the $\Gamma$ point was used
in the $k_x - k_y$ plane. The sampling along $k_z$ was
occasionally increased to 50 {\bf k}-points, when especially high
accuracy was needed. Stress along the wire direction, supported by
the periodic boundary conditions, is evaluated in our code
according to the Nielsen-Martin prescription\cite{Ole}, and
convergence was assumed when the stress along $z$ was relaxed at
around 10 kbar. Atomic relaxations were carried out using the
Hellmann-Feynman forces and convergence was assumed when forces
fell below $\sim$10 meV/\AA .

\section{Results}
\label{sec:results}

The calculated electronic structure for four among the seven wires
considered, with atoms in their fully relaxed positions, are shown
along with their density of states (DOS) in Fig.\ref{fig:bands}
a-d. All wires are metallic. Not surprisingly, the band structures
comprise a number of clear 1D subbands which disperse rather free
electron like along the wire. This band dispersion is strongly
reminiscent of quantized states in a cylinder; but of course only
roughly so, and the detailed atomic structure has a very
non-negligible impact. The DOS at $E_F$ jumps without continuity
from a wire to another. Needless to say, wires where this quantity
is lower will be comparatively more stable against instabilities.

In the ballistic transport regime, the number of bands crossing
the Fermi level determines the number of channels open for
conduction. The detailed bands do depend substantially upon the
underlying atomistic structure of the wire, suggesting that
jellium models should be inadequate to describe the detailed
physics of real nanowires.

The number of channels (as defined above) decreases with
decreasing wire diameter. If we introduce a  conventional {\em
wire radius} using, for example, Gulseren et al.'s
\cite{Gulseren98} definition, (a sort of local gyration radius)
then we can plot the channel number versus radius, as in
Fig.\ref{fig:channels}. In the same figure we compare our
predictions with a hard-wall cylinder jellium model, whose
solution are Bessel functions, and with {\em soft-wall} jellium
model \cite{Martin}, whose solutions implies just a rigid
translation of the stepped curve, which depends upon the Fermi
energy and the work function of the material (for Al we used $E_f
= 11.7$ eV and $\phi = 4.28$ eV).

If we note that the hexagonal wire (with 11 channels) is the
smallest of the possible {\em crystalline} fcc wires, we can
conclude that any observed conductance plateau which might be
traced back to less than 11 channels, is in itself indirect
evidence of a noncrystalline wire. In particular, pentagonal ({\em
icosahedral}) wires are predicted to possess  8 or 10 conducting
channels, in the eclipsed and the staggered pentagonal wires
respectively.

The energy per particle decreases for increasing radius, as
expected from size corrections of the type $E(R)$ = $E(\infty) +
2\gamma/R$, where $\gamma$ is the surface energy. Comparison with
the glue model energies is shown in Fig.\ref{fig:energy}. We note
that our calculations tend to predict somewhat higher  cohesive
energies. A fit to this energies, up to the two-fold wire,
predicts a surface energy of $\gamma = 45\pm 5 {\rm eV}/{\rm
\AA}^2$. The monoatomic wire is observed to lie out of this
fitting curve.

\section{Discussion}
\label{sec:discuss}

We have calculated the GGA electronic structure and relaxed total
energy of a sequence of ultra-thin aluminum nanowires. Despite its
resemblance with the corresponding jellium result, the detailed
electronic structure is in fact strongly dependent on the precise
atomistic structure. The number of conducting channels has been
determined, and it is found that channel numbers below 11 cannot
correspond to crystalline wires. In particular, the pentagonal
wires discovered theoretically by Gulseren et al. should possess 8
or 10 channels, implying an ideal ballistic conductivity of 8 or
10 times ($2e^2/h$). In future we will consider the eventual
transition to (anti-)ferromagnetic states of some of the wires,
allowing also for possible dimerizations. This transition might be
induced stretching the wire in order to {\em push} the Fermi level
near to a singularity of the DOS, as predicted with a jellium
model by Zabala {\em et al.}\cite{Zabala}.

We acknowledge support from  MURST COFIN97 and
by INFM (Sez. F e G). Work by F.D.D. was
carried out under an INFM/G fellowship ({\em Progetto Nanofili}).
Work by J.A.T. was carried out under TMR grant ERBFMBICT972563. We
would like to thank F. Ercolessi, G. Chiarotti, and S. Scandolo
for useful discussions.

\begin{figure}
\narrowtext

\caption{Band structures for four different Al nanowires, and
relative density of states: (a) monoatomic wire, (b) triangular
wire, (c) pentagonal staggered wire, and (d) hexagonal eclipsed
wire.}

\label{fig:bands}
\end{figure}

\begin{figure}
\narrowtext

\caption{Number of channels open for conduction. The dashed line
is the number of channels predicted for a jellium system confined
in a hard-wall cylinder, as computed in Ref. {\protect
\cite{Martin}}. The dashed line (soft-wall) is the prediction for
the same system corrected to better describe a real wire (see
text).}

\label{fig:channels}
\end{figure}

\begin{figure}
\narrowtext

\caption{Cohesive energy vs. the inverse of the wire radius. Open
symbols correspond to GGA calculations. Filled symbols to
classical MD from Ref. {\protect \cite{Gulseren98}}. Diamonds
correspond to hexagonal wires, squares to pentagonal, up-triangles
to triangular, down-triangles to two-fold wires, and the open
circle to monoatomic wire. The line is a fit to the formula $E(R)
= E(\infty) + 2\gamma/R$ and predicts $\gamma=45\pm 5$ meV/\AA$^2$
.}

\label{fig:energy}
\end{figure}


\begin{references}

\bibitem{Gulseren98}  O. Gulseren, F. Ercolessi and E. Tosatti, Phys. Rev. Lett.
{\bf 80} (1998) 3775.

\bibitem{Krans} J.M. Krans, C.J. Muller, I.K. Yanson, Th.C.M. Govaert, R. Hesper,
and J.M. van Ruitenbeek,  Phys. Rev. B {\bf 48}  (1993) 14721;
J.M. Krans, J.M. van Ruitenbeek, V.V. Fisun, I.K. Yanson, and L.J.
de Jongh, Nature {\bf 375}  (1995) 6534.

\bibitem{Agrait} N. Agra\"{\i}t, J.G. Rodrigo, and S. Vieira, Phys.
Rev. B {\bf 47}  (1993) 12345; N. Agra\"{\i}t, J.G. Rodrigo, C.
Sirvent, and S. Vieira, Phys. Rev. B {\bf 48}  (1993) 8499; N.
Agra\"{\i}t, G. Rubio, and S. Vieira, Phys. Rev. Lett. {\bf 74}
(1995) 3995.

\bibitem{Pascual} J. I. Pascual, J. M\'{e}ndez, J. G\'{o}mez-Herrero, A. M. Bar\'{o}, N. Garc\'{\i}a, and V.T.
Bihn, Phys. Rev. Lett. {\bf 71}  (1993) 1852; J. I. Pascual, J.
M\'{e}ndez, J. G\'{o}mez-Herrero, A. M. Bar\'{o}, N. Garc\'{\i}a,
U. Landman, W. D. Luedtke, E. N. Bogachek, and H. P. Cheng,
Science {\bf 267}  (1995) 1793.

\bibitem{Olsen94} L. Olsen, E. Laegsgaard, I. Stengsgaard, F.
Beschenbacher, J. Schi{\o}tz, K.W. Jacobsen, and J.N. N{\o}rskov,
Phys. Rev. Lett. {\bf 72}  (1994) 2251.

\bibitem{Torres} J.A. Torres, J.I. Pascual, and J.J. S\'{a}enz, Phys. Rev. B {\bf 49},
(1994) 16581; J.I. Pascual, J.A. Torres, and J.J. S\'{a}enz, Phys.
Rev. B {\bf 55} (1997) 16029.

\bibitem{Martin} A. Garc\'{\i}a-Mart\'{\i}n,  J. A. Torres, and J. J.
S\'{a}enz, Phys.  Rev. B {\bf 54}  (1996) 13448.

\bibitem{Stafford97} C.A. Stafford, D. Baeriswyl, and J. B\"{u}rki,
Phys. Rev. Lett. {\bf 79}  (1997) 2863.

\bibitem{Campers}  M. Brandbyge, K.W. Jacobsen and J.K. N\o rskov Phys.Rev.B
{\bf 55}  (1997) 2637; M. Brandbyge, M.R. S\o rensen and K.W.
Jacobsen Phys. Rev. B {\bf 56}  (1997) 14956; see also Ref.
\cite{Olsen94}.

\bibitem{Lang95} N.D. Lang, Phys. Rev. B {\bf 52} (1995) 5335.

\bibitem{Barnet97} R.N. Barnet and U. Landman, Nature {\bf 387}
(1997) 788.

\bibitem{Yannouleas98} C. Yannouleas, E.N. Bogachek, and U.
Landman, Phys. Rev. B {\bf 57},  (1998) 4872.

\bibitem{Kobayashi99} N. Kobayashi, M. Brandbyge, and M. Tsukada,
Jpn. J. Appl. Phys. {\bf 38},  (1999) 336.

\bibitem{Tomagnini} O. Tomagnini, F. Ercolessi, and E. Tosatti,  Surf. Sci. {\bf 287/288},
(1993) 1041, O. Tomagnini, F. Ercolessi, and E. Tosatti, unpublished data.

\bibitem{Kuipers93} L. Kuipers and J.W.M. Frenken, Phys. Rev. Lett. {\bf 70},  (1993) 3907.

\bibitem{Takayanagi}  H. Ohnishi, Y. Kondo, and K. Takayanagi, Nature {\bf 395},
(1998) 780; Y. Ohnishi, and Takayanagi, Phys. Rev. Lett. {\bf 79}
(1997) 3455.



\bibitem{Takeuchi89} N. Takeuchi, C.T. Chan, and K.M. Ho, Phys.
Rev. B {\bf 40} (1989) 1565.

\bibitem{Ercolessi88}  F. Ercolessi, M. Parrinello, and E.  Tosatti,  Phil.  Mag.  A {\bf 38}, (1988) 213.

\bibitem{Andreoni91} F.Ercolessi, W. Andreoni, and E. Tosatti, Phys. Rev. Lett. {\bf 66} (1991) 911.

\bibitem{Scheer} E. Scheer, P. Joyez, D. Esteve, C. Urbina,
and M.H. Devoret, Phys. Rev. Lett. {\bf 78}  (1997) 3535; E.
Scheer, N. Agra\"{\i}t, J.C. Cuevas, A.L. Yeyati, B. Ludoph, A.
Mart\'{\i}n-Rodero, G.R. Bollinger, J.M van Ruitenbeek, and C.
Urbina, Nature {\bf 394} (1998) 154.

\bibitem{Balicas} L. Balicas {\em et al.}, unpublished (1999).

\bibitem{Ole} O.H. Nielsen and R.M. Martin, Phys. Rev. Lett. {\bf
50} (1983) 697.

\bibitem{Becke} A. D. Becke, Phys.\ Rev.\ A {\bf 38},  (1988) 3098; J. P. Perdew, Phys.\ Rev.\ B
{\bf 33},  (1986) 8822.

\bibitem{Gonze}  R. Stumpf, X. Gonze, and M. Scheffler,
Research Report of the Fritz Haber Institute, April 1990.



\bibitem{Zabala} N.Zabala, M.J. Puska, and R.M. Nieminen, Phys.
Rev. Lett. {\bf 80} (1998) 3336.

\end{references}
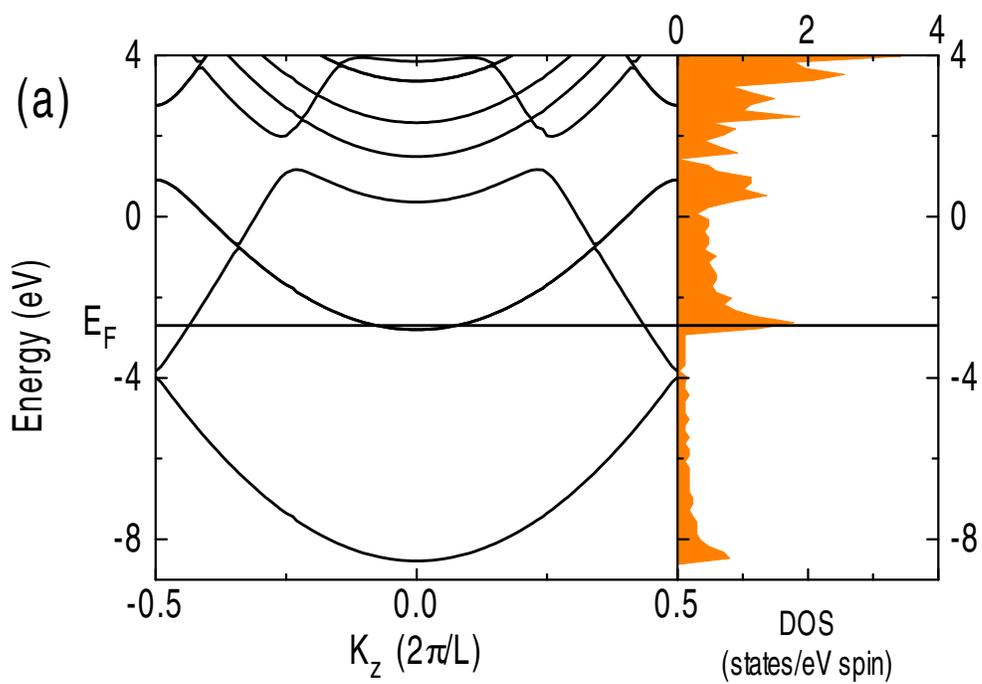
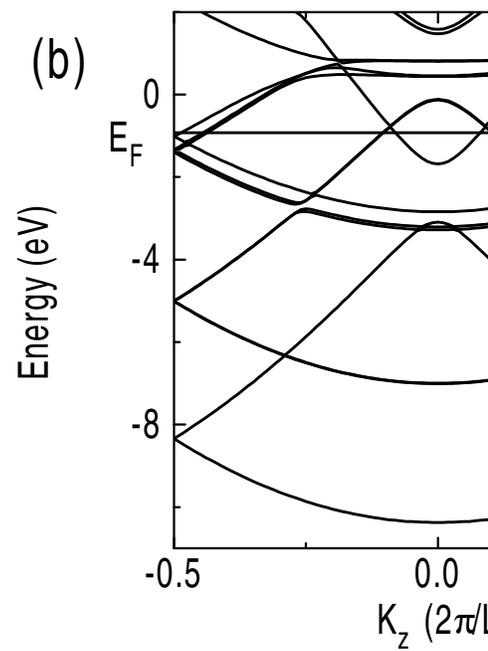
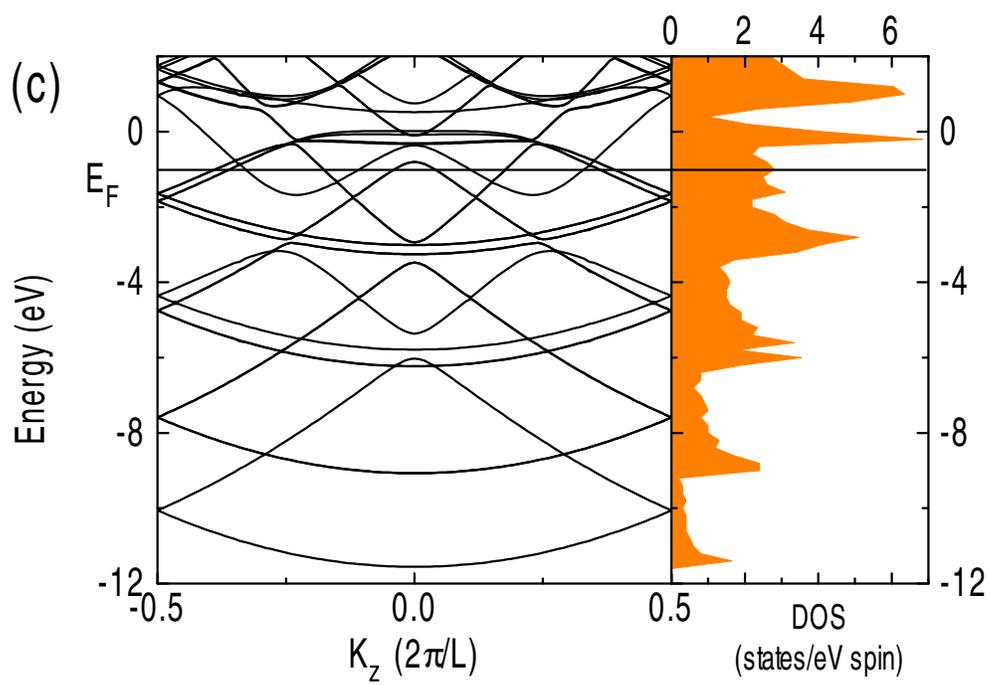
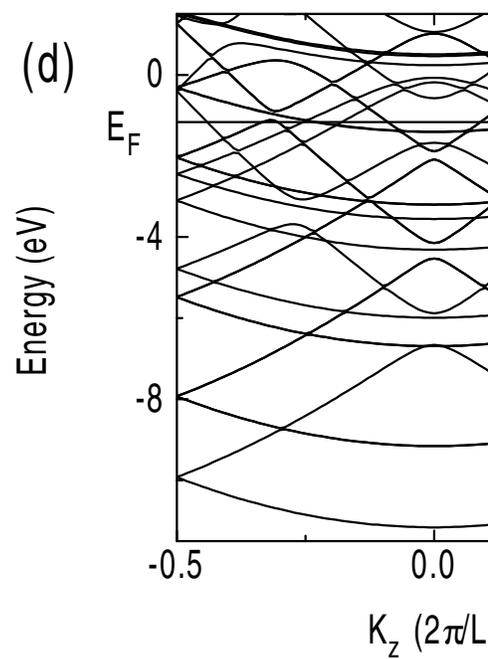
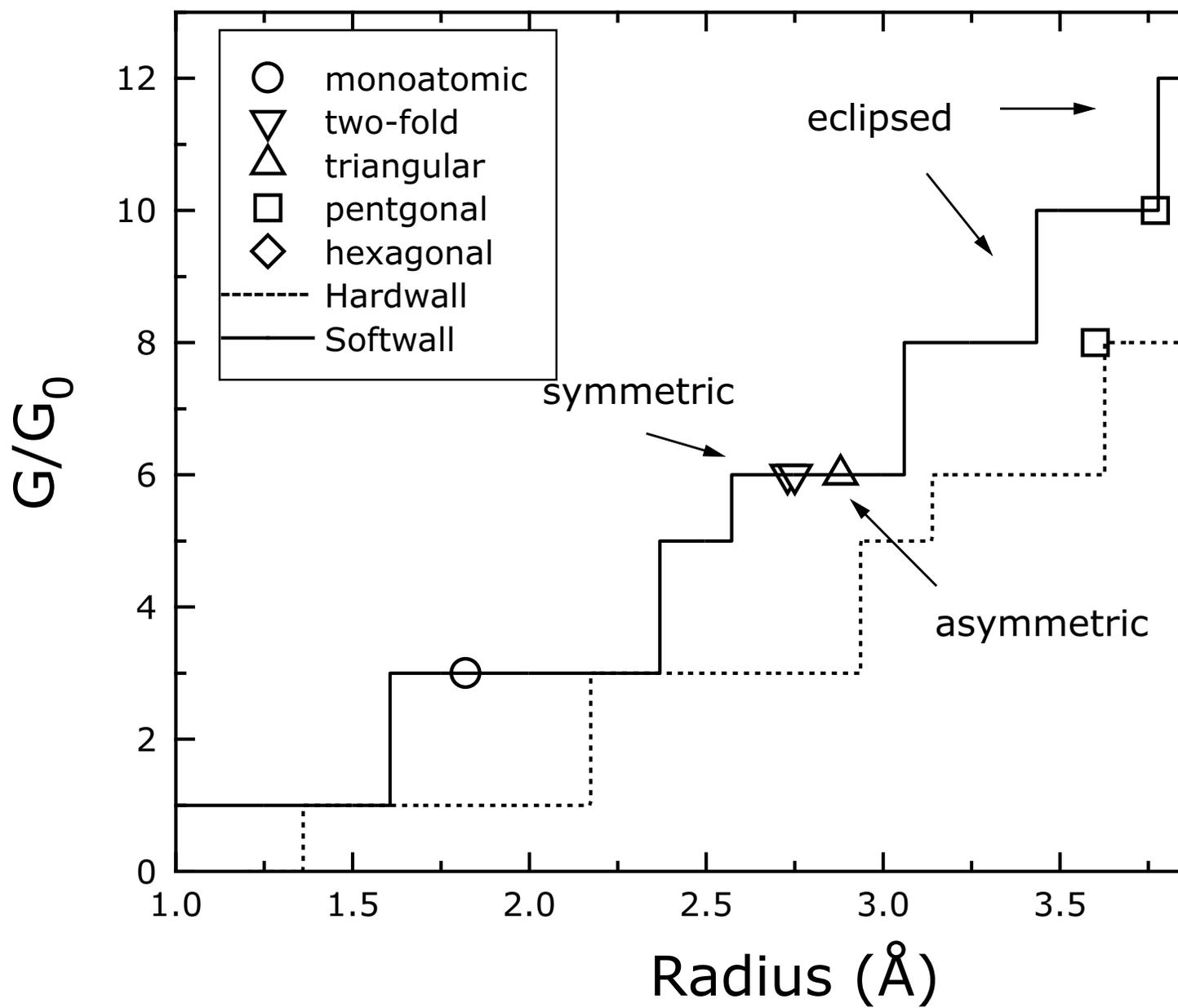
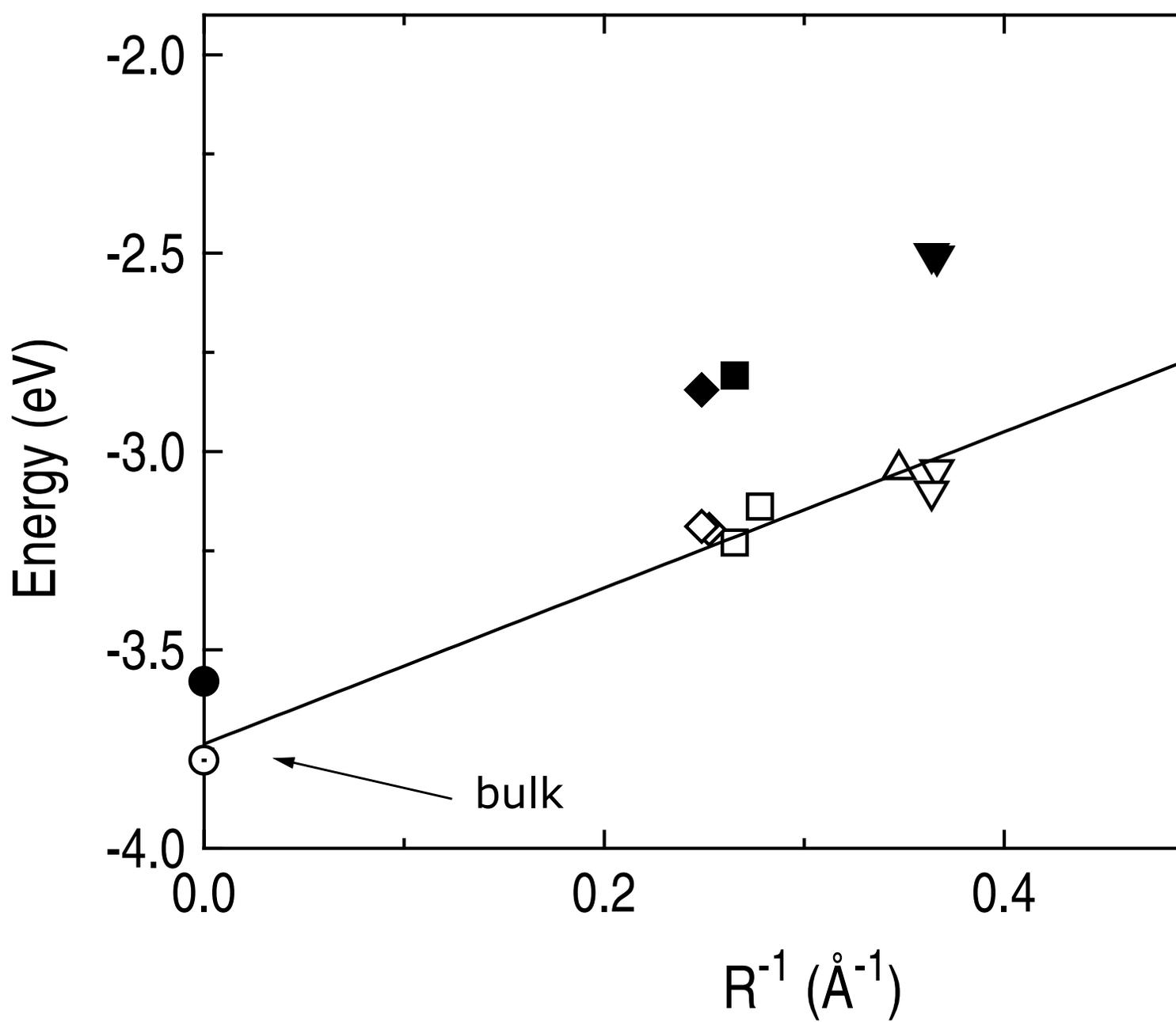
\end{document}